# Blockchain Technology: Bitcoins, Cryptocurrency and Applications

## Bosubabu Sambana

Assistant Professor,  Department of Computer Science and Engineering,
Raghu Engineering College (A), Visakhapatnam,
Jawaharlal Nehru Technological University Kakinada, Andhra Pradesh, India.
bosukalam@gmail.com

**Abstract —** *Blockchain is a decentralized ledger used to securely exchange digital currency, perform deals and transactions efficient manner, each user of the network has access to the latest copy of the encrypted ledger so that they can validate a new transaction. The blockchain ledger is a collection of all Bitcoin transactions executed in the past. Basically, it's a distributed database that maintains continuously growing tamper-proof data structure blocks that holds batches of individual transactions. The completed blocks are added in a linear and chronological order. Each block contains a timestamp and information link which points to a previous block. Bitcoin is a peer-to-peer permission-less network that allows every user to connect to the network and send new transactions to verify and create new blocks. Satoshi Nakamoto described the design of Bitcoin digital currency in his research paper posted to a cryptography listserv 2008.*

*Nakamoto's suggestion has solved the long-pending problem of cryptographers and laid the foundation stone for digital currency. This paper explains the concept of Bitcoins, its characteristics, the need for Blockchain, and how Bitcoin works. It attempts to highlight the role of Blockchain in shaping the future of banking, financial services, and the adoption of the Internet of Things(IoT), Artificial Intelligence and future Technologies.*

## 1. INTRODUCTION

The cost of cyber-crime costs quadrupled from 2013 to 2015 however a large portion of cybercrime goes undetected. Gartner report says cost of cyber-crime is expected to reach $2 trillion by 2019. IBM's CEO, Ginni Rometty said that cyber- crime is the greatest threat to every company in the world at IBM Security Summit. Around two years ago Standard Chartered lost around $250 million in a fraud at China's Qingdao port. Banking and financial institutions are using Blockchain based technology to reduce risk and prevent cyber fraud.

For example, Nasdaq has announced its plan to launch Blockchain based digital ledger technology which will help to boost their equity management capabilities. Standard Chartered is partnering with

DBS Group to develop an electronic invoice ledger using a Blockchain.

Blockchain can play crucial role in Internet of Things (IoT) and development of smart systems since we can track the history of individual devices by tracking a ledger of data exchanged. It can enable smart devices to act like an independent agent which can autonomously perform several transactions. For example, smart home appliances competing with one another for priority so that laundry machine, thermostats, dishwasher and smart lighting run at an appropriate time to minimize cost of electricity against current grid prices. Another example could be smart vehicles which can diagnose any problem and schedule to pay for its maintenance.

## 2. RELATED WORK

Blockchain is a transaction database which contains information about all the transactions ever executed in the past and works on Bitcoin protocol. It creates a digital ledger of transactions and allows all the participants on network to edit the ledger in a secured way which is shared over distributed network of the computers. For making any changes to the existing block of data, all the nodes present in the network run algorithms to evaluate, verify and match the transaction information with Blockchain history[1]. If majority of the nodes agree in favor of the transaction, then it is approved and a new block gets added to the existing chain.

The Blockchain metadata is stored in Google's LevelDB by Bitcoin Core client. We can visualize Blockchain as vertical stack having blocks kept on top of each other and the bottommost block acting as foundation of the stack. The individual blocks are linked to each other and refers to previous block in the chain [2].

The individual blocks are identified by a hash which is generated using secure hash algorithm (SHA-256) cryptographic hash algorithm on the header of the block. A block will have one parent but can have multiple child each referring to the same parent block hence contains same hash in the previous block hash field. Every block contains hash of parent block in its own header and the sequence of hashes linking individual block with their parent block creates a big chain pointing to the first block called as Genesis block.

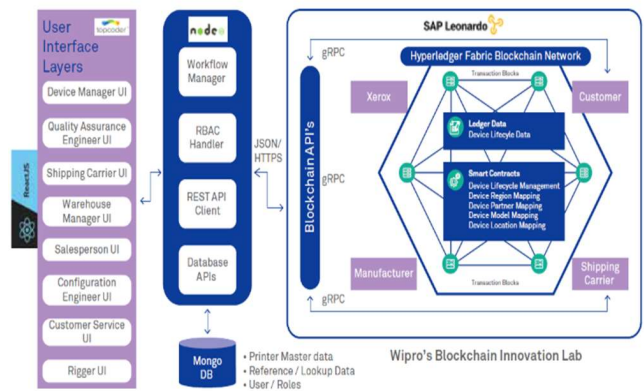

Figure.1: Blockchain enhanced digital transformation in enterprise architecture

### What is Bit Coin

Bitcoin is digital currency released as an open-source software in 2009. It is a decentralized cryptocurrency produced by all the participating nodes in the system at a defined rate. The chain of Bitcoins created over period and linked to each other called Blockchain [3]. It can be used to search any past transaction happened over the network between Bitcoin addresses. When a new block of

Bosubabu Sambana

transactions is created, it gets added to the Blockchain.

The new transaction records are continuously added to Bitcoin's public ledger and this process is called Bitcoin mining. Secure Hash Algorithm 2 (SHA-2) which is a cryptographic hash function is used by Bitcoin. We can determine integrity of a given data by comparing the execution output of SHA-2 algorithm called "hash" with an already known and expected hash value [4]. A hash algorithm converts a large volume amount of data into a fixed-length hash. And same data will always produce same hash but any slight modification in data will completely change the hash.

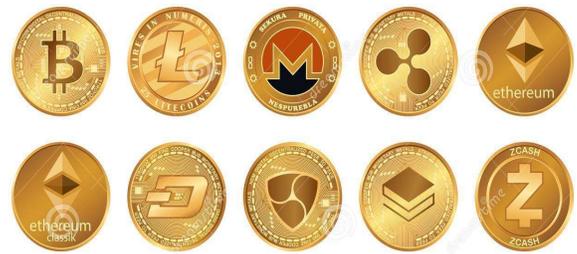

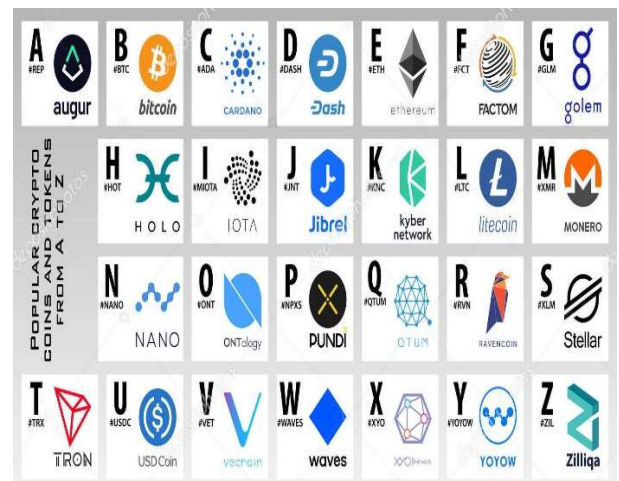

Figure.2: Types of Cryptocurrencies

### Genesis Block

The first block #0 created in 2009 is referred as Genesis block in Blockchain. It is common ancestral parent of all the new blocks created and if traversed backward in time we will reach genesis block in the end.

Genesis block is common ancestor of all the blocks and was created in 2009. It is encoded within bitcoin client software and can't be tampered.

All the node always knows the hash and structure of genesis block which is secure root. The statically encoded genesis block can be seen inside the Bitcoin Core client in chainparams.cpp. We can look for exact block hash:

"000000000019d6689c085ae165831e934ff763ae46a2a6c1 72b3f1b60a8ce26f'' in the block explorer websites to find details of Genesis block, latest transactions and all the newly created blocks with BlockHash, height, next block, size in bytes .

Genesis block contains a text message "The Times 03/Jan/2009 Chancellor on brink of second bailout for banks". The message was embedded in the first block by Bitcoin's creator Satoshi Nakamoto.

It offers proof of date when Genesis block was created which refers to headline of British newspaper The Times. The donation for Genesis block was made by John Wnuk and Jayden McAbee on June 9, 2016 which contains the first Bitcoin wallet.

### Prerequisites of Bit Coin

- **Authentication:** BitID which is a decentralized authentication protocol allows users to connect with Bitcoin. BitID uses Bitcoin wallets and QR codes to provide service or platform access points.
- **Integrity:** Bitcoin's use of digital signatures ensures transactional integrity and transactions can't be modified later.
- **Non-Repudiation:** The person who sent the message had to be in possession of the private key and therefore owns the Bitcoins. The sender needs to sign the previous hash and the destination public key.

### Benefits of Bitcoins

- **Fast & Cheaper:** The transactions made using Bitcoin's wallets are fast and transaction fees are minimal.
- **Decentralized Registry:** Bitcoin currency is decentralized and no central authority has full control and hence central government or banks can't take it away from you and there is no chargeback. The Central Bank during financial crisis in Cyprus wanted to take back all uninsured deposits more than $100,000 in 2013 to help recapitalize itself . But this is not possible with Bitcoins since currency is decentralized.
- **Secure Payment Information:** Bitcoin transactions use a public key and a private key. When a bitcoin is sent, the transaction is signed by public and private keys together which creates a certificate.
- **Bitcoin Mining:** You can create your own money by setting up a Bitcoin Miner

### How Bitcoins works

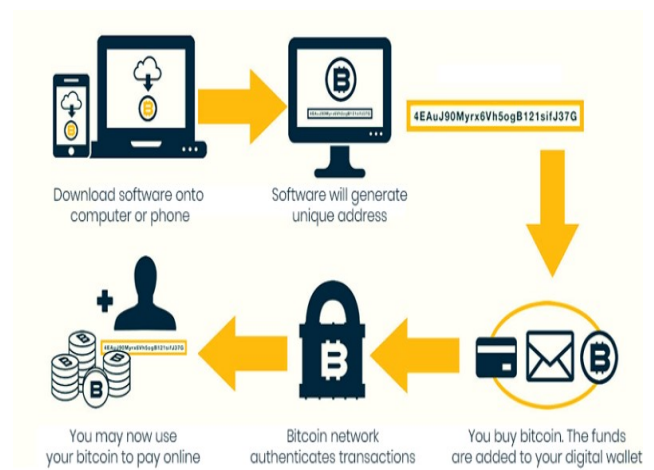

Figure.3: How do Bitcoins works

Bitcoin uses Elliptic Curve Digital Signature Algorithm(ECDSA) cryptographic algorithm to make sure only rightful owners have the access to funds. When Bitcoin is sent, it creates a transaction message and attaches new owner's public ECDSA key. Each Bitcoin is associated with public ECDSA

Bosubabu Sambana

key of its current owner[5]. A new transaction is broadcasted over Bitcoin network to inform everyone that new owner of these coins is the owner of the new key.

Bitcoin kiosks are machines which are connected to the internet and allow to deposit cash in exchange of Bitcoin's given as a paper receipt or by moving money to a public key on the Blockchain. Whenever a Bitcoin is sent, it attaches the new owner's public key and sign it with the sender's private key.

The sender's signature on the message verifies that the message is authentic and transaction history is kept by everyone so it can be easily verified. It uses Public Key Cryptography Asymmetric Encryption algorithm and concept of public and private keys to encrypt and decrypt data. If message is encrypted using public key($P_k$), then private or secret key($S_k$) is necessary to decrypt. However, when message is encrypted using private or secret key($S_k$) then public key($P_k$) is necessary to decrypt [7].

The public key can be shared with anyone but private key needs to be kept secret. One participant can create multiple public key(Pk) and secret key(Sk) pair. Bitcoin does not require third party as it publicly distributes the ledger called Blockchain. The interested users in devoting CPU power to run a special piece of software are called Bitcoin miners and they ultimately form a network to maintain the Blockchain [8]. In the Bitcoin mining process, users create new Bitcoin currency and transaction is broadcasted over the network [9].

All the computers running the software in network compete to solve cryptographic puzzles which contains data from several transactions. The first miner who solves each puzzle get 50 new Bitcoins and the corresponding block of transactions gets added to the existing Blockchain. The difficulty level of each puzzle is directly proportional to the number of Bitcoin miners present [11]. As the number of miners increases, the difficulty level of puzzle is also increased to ensure production of one block of transactions for every 10 minutes.

Bosubabu Sambana

Bitcoin works on decreasing supply algorithm which means the reward for mining Bitcoin block is reduced to half after every 210,000 blocks. The block creation rate is adjusted every 2016 blocks to ensure creation of roughly 6 blocks per hour. The number of Bitcoins generated per block is set to decrease geometrically and it will reduce by 50% every 210,000 blocks which is approximately 04 years' time. The aim is to ensure number of Bitcoins in existence never exceeds 21 million since monetary base of Bitcoins cannot be expanded therefore Bitcoin currency would undergo severe deflation if it becomes widely used [12].

### Bitcoin Advanced standards

Bitcoin transactions are made using script which is stack-based and processed from left to right. The script contains list of instructions recorded with each transaction having description on how receiver can gain access to it. The transactions are valid if script returns true.

```
if
verify_signature(
transaction.signature,
transaction.input.public_key):
return True
else
return False
```

The standard transactions on Bitcoin network are called single-signature transactions because it requires just one signature from the owner. But Bitcoin also supports complex transactions which require signature of more than one people.

It supports Multisignature (Multisig) transaction which sends funds from a multi-signature address and referred as M-of-N transactions which is associated with N private keys and requires signatures from at least M keys Lock times is a bitcoin feature which makes a transaction not allowed to the network until a certain time [13].

The locked transaction will spend the coins it uses as inputs at a certain time in the future, unless those coins are first spent by a prior transaction.

Micropayment channels use both multi-sig technology and a lock time. Bitcoin multisig wallets have potential for increasing the security of funds and giving technology tools to enforce better corporate governance. Each transaction is added to the blockchain after consensus among nodes and verification to protect against double spending. The transactions are validated when a mathematical problem is solved [14].

The Bitcoin transactions require a time window before they are confirmed which is an issue since speedy transaction is expected from digital currency in the modern world. But using marker address also called as green address approach we can reduce the time taken for confirmation of Bitcoin transaction. The marker addresses use Bitcoin's own communication channel and parties can establish trust sine a receiver is already waiting for an address to send money [15].

Pay to script hash (P2SH) transactions is supported by Bitcoin which allow transactions to be sent to a script hash instead of a public key hash. The recipient must provide a script matching to the script hash to spend bitcoins sent via P2SH and the data which makes script evaluation true.

We can send bitcoins using P2SH to an address that is secured. The bitcoins can be sent to ~34-character P2SH address and recipient needs signatures of several people to spend bitcoins or a password or there could be entirely unique requirements. Byzantine consensus problem is used in a large peer-to-peer network ·

All the processes in the network should come to a common agreement. The networks without admission controls are vulnerable to the Sybil attack in which malicious processes can claim multiple fraud identities[15].

Satoshi Nakamoto inventor of the Bitcoin protocol published his research paper called "Bitcoin: A Peer-to-Peer Electronic Cash System" in 2008. This paper talked about peer- to-peer electronic cash transfer directly from one party to another without going through the channel of financial institutions

Ethereum is a public Blockchain based distributed computing platform which runs smart applications without any downtime or third party interference. These applications run on a custom built Blockchain which enables to store registries of debts or promises and move funds without requiring a middle man or counterparty risk [16].

Ethereum provides a decentralized virtual machine called Ethereum Virtual Machine (EVM) which can execute peer-to-peer contracts using a cryptocurrency called Ether. Ethereum project was bootstrapped by an Ether pre-sale during 2014 and developed by a Swiss nonprofit group called Ethereum Foundation. Blockchain based DNS and Blockchain based internet could be future.

The DNS Chain offers a free and secure decentralized alternative while remaining backwards compatible with traditional DNS.

### Blockchain adoption Trends

Price Waterhouse Coopers (PwC) reports show Blockchain is being explored and adopted at an unprecedented speed. Blockchain is still five to ten years from mainstream adoption but it's already reached peak of inflated expectations reported by Gartner in Hype Cycle for Emerging Technologies 2016 Blockchain stores multiple transactions in one centralized ledger which is accessible to all parties and activities are regulated by a decentralized network. There are around dozens of offerings on distributed-ledger products in the market but Bitcoin is most popular and proven among all.

The 3 trends appearing in Gartner Hype Cycle for emerging technologies 2016 are Blockchain, Smart machines giving transparent immersive experiences like 4D printing and Internet of Things (IoT)

Bosubabu SambanaarXiv:2107.07964v1 [cs.CR] 16 Jul 2021

leading to perceptual smart machine transparency age in land registry. They are using Blockchain technologies to update and maintain land registry records to prevent corruption since once land records are verified it can't be tampered.

IBM and Samsung are working on a proof of concept (PoC) to build next generation Internet of Things(IoT) which will be based on Autonomous Decentralized Peer-to-Peer Telemetry(ADEPT) which uses BitTorrent peer-to-peer file sharing protocol, Rehash peer-to-peer communication protocol, Bitcoin cryptocurrency & Ethereium.

Everledger is using Blockchain to tackle conflict diamonds and insurance fraud since tracking of diamonds from mine to retail stores is a lengthy and complex process. They are combining data from insurance companies and police departments to provide an accurate database which can be made available to everyone on Blockchain [17].

Distributed Autonomous Organization(DAO) is an organization meant to support Ethereum-related projects. DAO has received over $50m worth of digital token called ethers (ETH) from investors. People supporting DAO receive voting rights in form of digital token to determine future direction of the organization and support startups and projects by dispensing ethers (ETH). The participants in the voting and decision making process receive dividends for supporting the project.

Nasdaq has announced plan to launch Blockchain-enabled digital ledger technology which will help to expand and enhance its equity management capabilities. They will initially leverage Open Assets Protocol which is a Bitcoin based colored coins' implementation.

Bitfinex is Hong Kong based world's leading Bitcoin exchange which one of the most advanced cryptocurrencies exchange offering margin trading, advanced order types & margin funding for low-risk returns.

Bosubabu Sambana

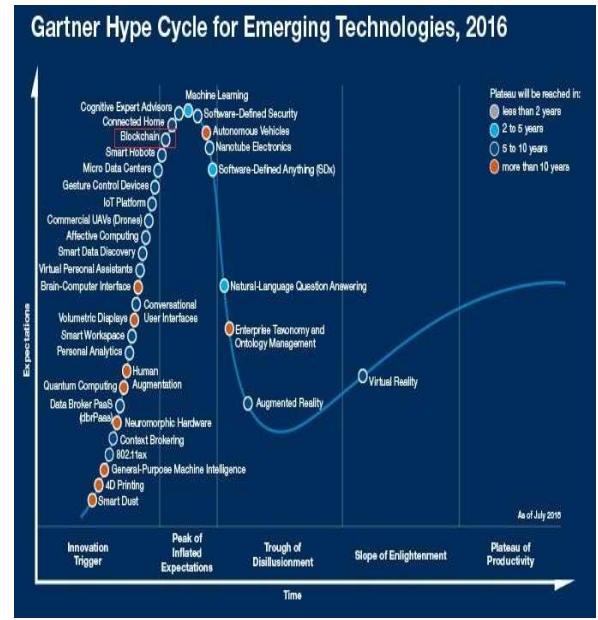

Figure. 4: Gartner Hype Cycle 2016

IBM Blockchain service on Bluemix provides four-node development and test blockchain network to write applications and deploy chain code immediately instead of creating Blockchain network from the scratch. It is basically a peer-to-peer permissioned network which is built on top of Hyperledger Fabric from Hyperledger Project of Linux Foundation.

The land registry system is broken in Honduras, a Central American country due to corruption. Government of Honduras announced a deal with Factom Inc, a Texas-based Blockchain company to implement Factom's Land Registry Tool to create

### 3. SUMMARY AND CONCLUSION

Venture capital(VC) firms are betting big on Bitcoin and the Blockchain. CoinDesk's Bitcoin Venture Capital data shows that VC firms are investing heavily in bitcoin startup projects. Blockchain can act as ledgers or record-keeper for billions of transactions generated by Internet of Things (IoT) since sharing, storing data and information is always a risk. Mckinsey report says that Blockchain have potential to reshape the capital markets industry with impact on business models, reductions in risks, cost and capital savings. The adoption of Blockchain technology will have significant benefits and reduce number of ledgers required to be maintained by financial institutions and ensure more precise audit trails.

A recent study from IBM Institute for Business Value (IBV) shows that 70 percent of early adopters surveyed are prioritizing Blockchain efforts to reduce barriers in creating new business models to reach new markets. The seven out of 10 early adopters surveyed in financial market institutions have Blockchain efforts focused mainly on four areas: clearing and settlement, equity and debt issuance, wholesale payments and reference data. The device cost is decreasing and computing power is increasing every day therefore Blockchain presents an immense possibility in Internet of Things(IoT) and providing security. Blockchain can offer a trustless system having peer- to-peer messaging protocols and secured distributed data sharing.

Bosubabu Sambana